\documentclass[twocolumn,prl,nobalancelastpage,aps,10pt]{revtex4-1}
\usepackage{graphicx,bm,times}
\usepackage{textcomp}
\usepackage{natbib} 
\usepackage{etoolbox}
\usepackage{amsmath}
\usepackage{amssymb}
\usepackage{subfigure}

\usepackage[margin=13.6mm,footskip=0.mm]{geometry}

\begin{document}

\title{\vspace{-5.0cm}Parallel N-body simulations of planetary systems: a direct approach}

\author{Dhananjay Saikumar}
\affiliation{H.H. Wills Physics Laboratory, University of Bristol, Tyndall Avenue, Bristol BS8 1TL, UK.}

\renewcommand{\thesubsection}{(\alph{subsection})}

\begin{abstract} 
Direct gravitational simulations of n-body systems have a time complexity $\mathcal{O}({n^2})$, which gets computationally expensive as the number of bodies increases. Distributing this workload to multiple cores significantly speeds up the computation and is the fundamental principle behind parallel computing. This project involves simulating (evolving) our solar system for the next 1000 years (from 2015 to 3015) on the BlueCrystal supercomputer. The gravitational bodies (planets, moons, asteroids) were successfully simulated, and the initial states (mass, position and velocity vectors) of the solar system objects were obtained via NASA's JPL Horizons web interface. Two parallel computing domains are investigated: shared and distributed memory systems.  
\end{abstract}

\date{{February 2021}}
\maketitle

\section{1. Introduction}
The transistor plays a central role in computing, and at its core, the transistor is a binary switch that can regulate the current flow. Transistors can be arranged differently to create logic gates and perform basic arithmetic operations. In the 50s, primitive computers transitioned from vacuum tube transistors to discrete transistors (wired on a circuit board). As the computing demand grew, wiring many individual transistors together set a physical limit to the density of transistors on a circuit board. The 60s brought the development of the metal-oxide-silicon (MOS) integrated circuit (IC), which used the process of chemical etching to connect the circuit components (significantly reducing the transistor-circuit size), which soon became the new standard \cite{1}. The MOS paradigm led to the development of the microprocessor, an IC which can process data according to instructions (serial) stored in its memory \cite{2}. 

The Modern CPU is a single IC which contains multiple microprocessors (multiple cores), where each core can run separate instructions simultaneously. The memory architecture of a generic multi-core processor is as follows: Starting at the smallest level, every core has a small pool of fast memory exclusive to itself (L1 cache and a larger, slower L2 cache). Going a level above, the cores share a larger but relatively slower pool of memory (L3 cache), allowing individual cores to share data (see Figure  1). Furthermore, individual CPUs can be linked together via a low latency interconnected network to distribute large amounts of computing workload.

\setlength\belowcaptionskip{0ex}
\begin{figure}[h]
\includegraphics[width=0.6\columnwidth]{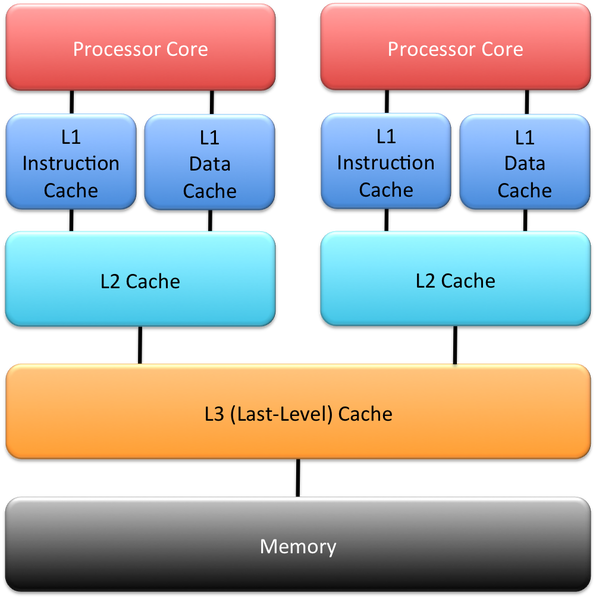}
\caption{Memory architecture in a modern CPU processor. Ref \cite{3}.}
\end{figure}
A serial algorithm contains a set of instructions computed sequentially. In contrast, parallelisation is a technique which involves breaking down the serial algorithm into smaller bits, and computing them in parallel on multiple cores, effectively speeding up the computation \cite{4}. The speedup obtained by a parallel algorithm running on $p$ cores is defined as:
\begin{equation}
S_{p} = \frac{T_{1}}{T_{p}} 
\end{equation} 
Where $T_{1}$ and $T_{p}$ are the time, the algorithm takes to run on a single core and p cores, respectively. Many parallel algorithms require synchronisation/communication between different distributed tasks. A crucial aspect of writing parallel code is recognising which tasks need to be communicated and which don't. Meanwhile, `embarrassingly parallel algorithms' (rarely need to communicate) are the easiest to parallelise.
Intuitively, a perfect program would speed up linearly with $p$. But no algorithm in the real world is perfectly optimal. The speedup a parallel algorithm can achieve is given by Amdahl's law \cite{5}:
\begin{equation}
T_{p} = T_{1} \left ( F_{s} + {\frac{F_{p}}{p}} \right )
\end{equation}
Where $T_{p}$ and $T_{1}$ are the same as before, $F_{p}$ and $F_{s}$ are the fraction of the algorithm, which is parallelizable and sequential, respectively. Ultimately, the $F_{s}$ sets the limit to the maximum speedup obtainable by an algorithm. For example, if 25\% of the algorithm is sequential, the maximum speedup achievable is 400\%, independent of ($p \geqslant p_{o}$) for some $p_{o}$, which depends on the algorithm.

\section{2. Direct N-body simulation}
This work aims to parallelize the N-body simulation of planetary systems. The simulation contains around 5000 solar system bodies (8 planets + Pluto, 191 natural satellites, 280 asteroid belt objects and 4500 other miscellaneous objects). N-body simulations are extensively used in astrophysics to study dynamical systems such as galaxy mergers, the evolution of star clusters, dark matter filaments, and more \cite{6}. In this simulation, the bodies are assumed to be point particles. The total force exerted on a particle ($i$) by other particles is described by Newton's law of universal gravitation:
\begin{equation}
\vec{F_{i}} = \sum_{j \neq i}\vec{F}_{ij} = \sum_{j \neq i}\frac{Gm_{i}m_{j}}{(\vec{r_{i}}-\vec{r_{j}})^{2}}\hat{r}
\end{equation} 
Where $F_{i,j}$ is the force exerted by the j$^{th}$ particle, $m_{j}$ and $r_{j}$ is the mass and position vector of the j$^{th}$ particle, respectively. This calculation is repeated for every particle in the system; since there are n particles in total, the time complexity of this computation is $O(n^{2})$. The force calculation is broken down into 3 parts (calculating forces in every direction ($\hat{x},\hat{y},\hat{z}$)). The general structure of the parallel algorithms used here is as follows:
 
The master core holds the current state (predefined position and velocity vectors) in memory. Every slave core is associated with a unique set of particles. The force is computed in parallel (each slave core calculates the force exerted on each of their particle by all particles in the system, in-turn calculating their  acceleration ($\hat{a_{x}},\hat{a_{y}},\hat{a_{z}}$)). The master core uses this (n$\times$3) acceleration matrix to update the previous state of the system via the leap-frog technique (kick-drift-kick) \cite{6}.
\begin{equation}
\vec{v}_{i+1/2} = \vec{v}_{i} + \vec{a}_{i}\frac{dt}{2}
\end{equation} 
\vspace{-5mm}
\begin{equation}
\vec{r}_{i+1} = \vec{r}_{i} + \vec{v}_{i+1/2}dt
\end{equation} 
\vspace{-5mm}
\begin{equation}
\vec{v}_{i+1} = \vec{v}_{i+1/2} + \vec{a}_{i+1}\frac{dt}{2}
\end{equation} 
Where $i$ is the current iteration label, $\vec{v}_{i+1}$ and $\vec{r}_{i+1}$ is the updated position and velocity vector, and $dt$ is the discrete time-step. The number of calculations can be reduced by half by utilising Newton's third law $\vec{F}_{ij} = -\vec{F}_{jj}$, although this requires an additional communication/synchronisation step.
The calculations discussed above (eq 4,5 and 6) are iterated $K$ number of times where ($K \times dt$ = total time simulated). A smaller time-step offers better accuracy but increases number of iterations (computationally expensive).

For a general N-body simulation, as the system evolves, the distance between two objects may approach 0, which generates a singularity in eq (3) (infinite acceleration). However, this isn't realistic since particles are finite sized. The singularity is avoided by introducing a softening term $\epsilon$ to the denominator of eq (3). This value needs to be small to keep the simulation realistic ($\epsilon \ll r$) \cite{7}. The softening term plays a crucial role when simulating gas clouds, galactic mergers which involve colliding particles. However, the softening term does not significantly impact the simulations of planetary systems because these systems tend to have stable orbits (no collisions). The parallel algorithm here is implemented separately via  
shared memory and distributed memory systems.

\subsection{2.1 Shared memory: OpenMP}
OpenMP (Open Multiprocessing) is an example of a shared memory interface. It uses the fact that cores on the same processor share a layer of memory (L3 cache), which eliminates the need for the cores to communicate (unlike MPI, see later). This reduces memory usage since only a single copy of the state needs to be kept in memory.
The OpenMP code initially runs on a master thread. The force is calculated in parallel. Once this is done, each core updates the shared memory layer with its piece of calculation, and then the serial code (running on the master thread) updates the state of the system via eq (4,5,6). Furthermore, OpenMP allows for dynamic load balancing (cores that finish early can be assigned more work), improving performance. A significant drawback of using OpenMP is that it is limited to a single processor (cache memory is not shared across multiple processors). 
\subsection{2.2 Distributed memory: MPI}
MPI (Message Passing Interface protocol) is an example of a distributed memory interface. The MPI code is designed to run on multiple cores which don't share a layer of memory. In order to transfer data, the cores must communicate with each other. This independence from a shared memory architecture allows MPI to scale with the total number of cores in the system (a cluster of interconnected processors). 

Unlike OpenMP, the MPI code runs on all the cores simultaneously. Tasks are split according to the rank of each core. A typical MPI hierarchy consists of a master core (rank 0) and slave cores. In this simulation, the master core distributes an equal amount of force calculations (first communication step) to every slave core. Once this is done, the slave cores send their calculations to the master core (second communication step). The master core then updates the system's previous state via eq (4,5,6). Depending on where the data is temporarily stored, cores assigned the same amount of work may not finish simultaneously.

Communication is a crucial aspect of MPI since sending and collecting data from the slave cores is necessary. Communication also causes overheads: Although communication latency may be low, recipient cores may have to wait occasionally since the data has to be packaged before being sent, resulting in inefficiencies. This algorithm sends and receives data using the MPI.Send and MPI.Recv commands require communication to take place before starting/resuming any work (blocking communication), causing inefficiencies. Non-blocking communication may result in race conditions (multiple cores may try to overwrite the same piece of data). 
Due to its shared memory architecture, OpenMP is expected to perform better than MPI when evaluated on a single processor. However, MPI is expected to offer better performance when evaluated on multiple processors (a large number of cores).
\subsection{2.3 Animation}
2D animations of 3D simulations were generated (easier to visualise). The initial states (mass, position and velocity vectors) of the aforementioned solar system objects were obtained via NASA's JPL Horizons web interface. The simulation was computed on BlueCrystal and was `evolved' for 1000 years. Figure  2 shows the state of the solar system on 01/01/2015 (start date) and 21/12/2020, where the great conjecture (alignment of Jupiter, Saturn and Earth) can be seen.

\begin{figure*}
  \includegraphics[width=\textwidth]{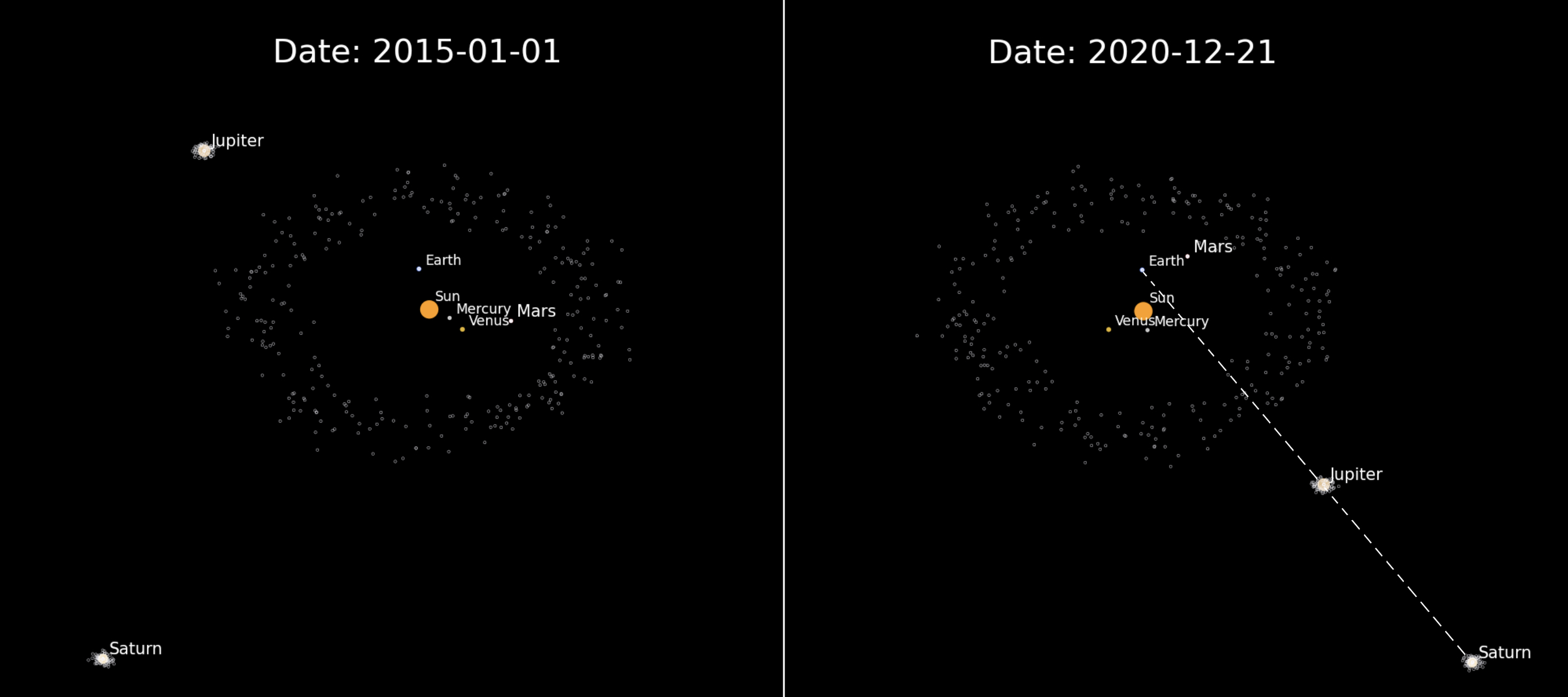}
  \caption{A snapshot of the animation at two different times. This simulation contains 5000 solar system objects. The main asteroid belt surrounds the inner planets. The Jupiter and Saturn systems include 67 and 80 natural satellites, respectively. Uranus and Neptune have been cropped out because their large orbits would drastically shrink the image. The Great conjecture of 2020 can be seen on the right. }
\end{figure*}
\section{3. Performance}
Different variations of the algorithm: OpenMP (cython), MPI (python), and serial (python) were implemented and timed on the BlueCrystal supercomputer. The speed up ($S_{p}$) from eq (1) is used as a measure of the algorithm's performance. Every measurement was repeated five times, and the mean and the standard deviation were used to generate the error bars. The spread in measurement is inherently unavoidable since it arises from background processes running on the same processor, potential overheads in accessing memory etc. The unit of time in the plots is arbitrary.
\subsection{3.1 Scaling: Model size ($n$) }
Figure  3 depicts how the performance of the algorithms scale with the model size ($n$). The algorithms were executed on a single BlueCrystal processor (16 $\times$ 2.6 GHz SandyBridge cores), using all 16 cores. For very small models ($n < 50$), the calculations are extremely small such that the serial (single core) algorithm performs better than MPI (calculations in serial are faster than distributing the calculations in MPI). Serial code also outperforms OpenMP for small models because the parallel cores can only access the shared cache memory one at a time (resulting in wait time). Additionally, the serial code (running on a single core) uses the fastest type of memory (L1 cache: $O(1000)$ GB/s), whereas the OpenMP code uses the shared memory layer, which is slower (L3 cache: $O(100)$ GB/s).
As $n$ increases, both parallel algorithms offer significant speedup over the serial algorithm ($S_{p} =$ 10 and 8.3 for OpenMP and MPI). As expected, OpenMP outperforms MPI due to the lack of communication overheads. It should be mentioned that math operations in cython (OpenMP) are much faster than python (MPI). Furthermore, `for loops' are much faster in cython due to the predefined variable type.

\subsection{3.2 Scaling: Number of cores ($p$)}
Figure  4 represents the performance of both parallel algorithms as a function of the number of cores $p$ for a fixed model size ($n =$ 5000). Here the OpenMP algorithm was timed on a single BlueCrystal processor, and the MPI algorithm was timed on two processors, utilising a total of 32 cores.

\setlength\belowcaptionskip{-2ex}
\begin{figure}
\includegraphics[width=0.8\columnwidth]{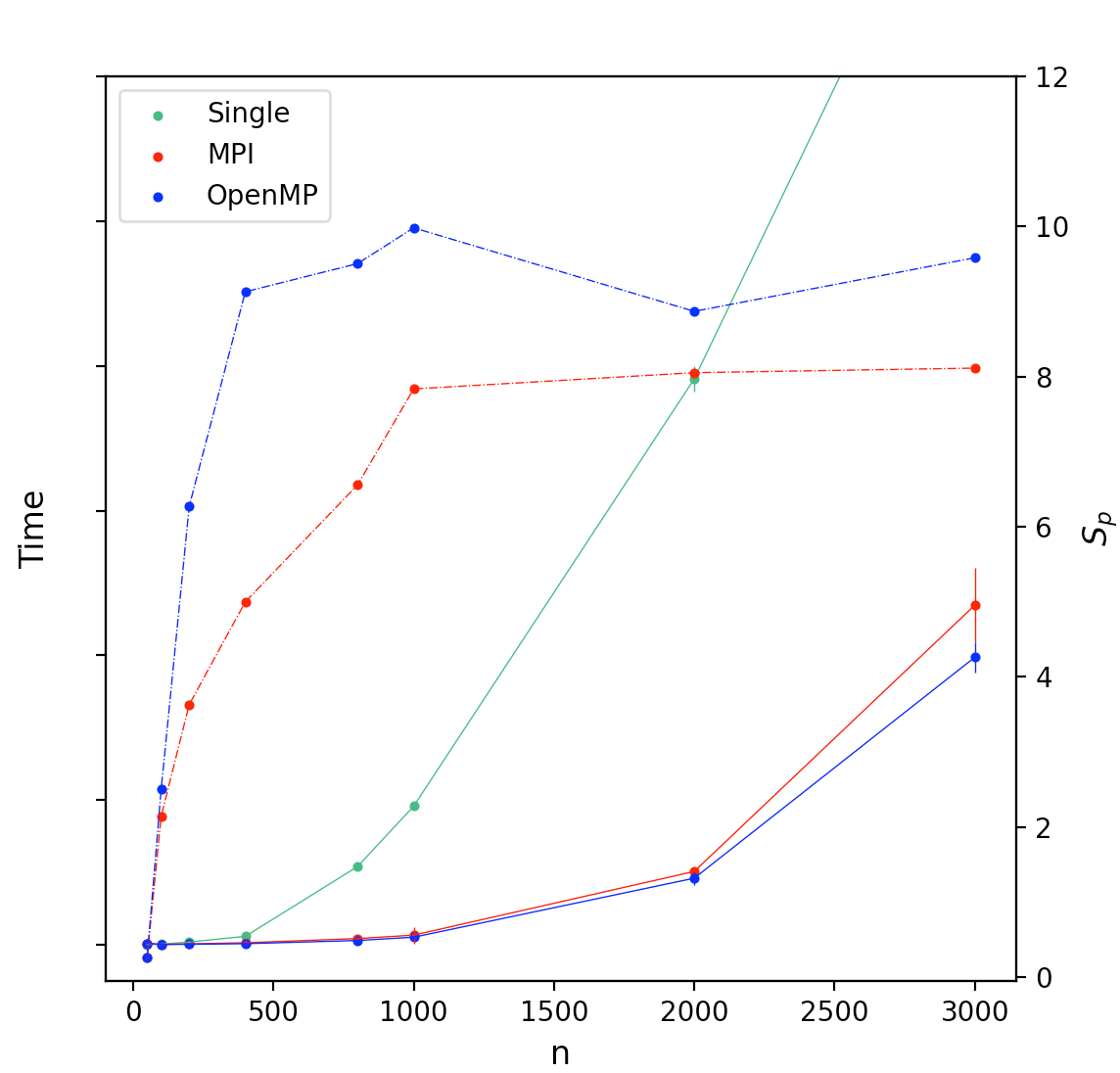}
\caption{Comparing how the algorithms scale with the model size ($n$). All algorithms were executed on a single BlueCrustal node (16 cores). The solid line represents the elapsed time (all algorithms show the $n^{2}$ time complexity), and the dotted line represents the speedup.}
\end{figure}

Figure  5 shows the speedup obtained by the MPI code for a range of cores ($2\leq p \leq32$) and multiple model sizes ($1000 \leq n \leq 15000$). The performance of each model size scales differently with $p$. Smaller models ($n \leq 2000$) require smaller calculations, which can be calculated optimally on a small number of cores. Distributing this calculation to many cores takes additional (extra communication) time, reducing the performance. 

Larger models ($5000 \leq n \leq 15000$) scale almost linearly with the number of cores. Although, it takes longer to package and distribute huge models, which is why ($n = 15000$) performs slightly worse than ($n = 5000, 10000$).
Although not plotted here, the efficiency of the algorithm defined as ($S_{p}/p$) gives an insight into how well the algorithm utilises all the cores assigned to it. As the speedup stagnates for model sizes ($n = 1000, 2000$), the algorithm's efficiency decreases with increasing $p$. 

\begin{figure}[h]
\includegraphics[width=0.8\columnwidth]{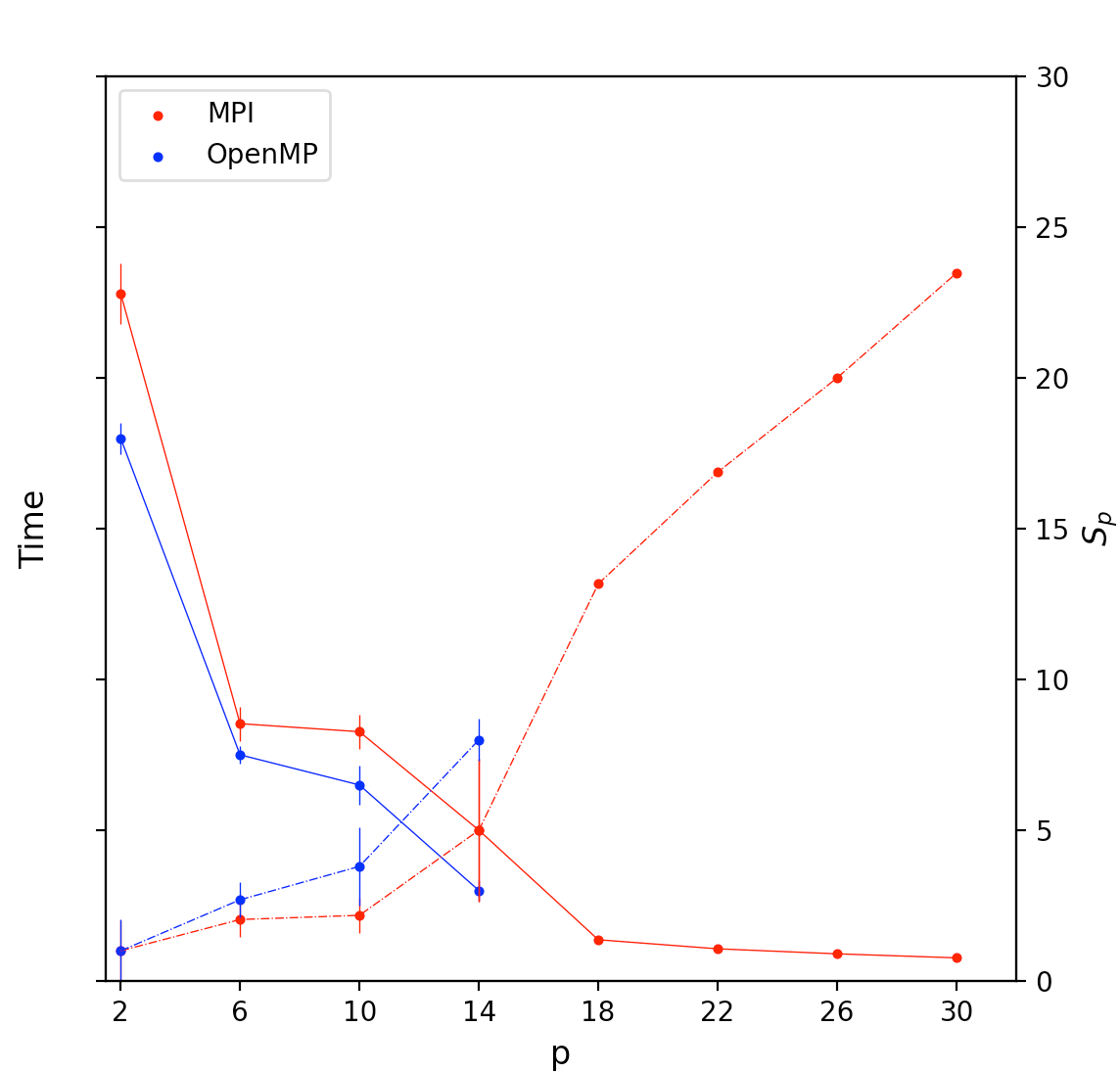}
\caption{Comparing how the algorithms scale with the number of cores ($p$), for constant model size ($n =5000$). The OpenMP code was executed on a single BlueCrustal node (16 cores), and the MPI code was run on two BlueCrustal nodes. The solid line represents the elapsed time, and the dotted line represents the speedup. As expected OpenMP outperforms MPI on a single processor ($p \leq 16$). But MPI outperforms OpenMP when evaluated on a larger number of cores.}
\end{figure}

Load balancing is the largest source of overheads in both MPI and OpenMP algorithms. Some cores may finish their calculations faster due to: memory latency (depending on where the cache is stored temporarily) and background interference. This results in some cores finishing up faster, staying idle till all the cores have caught up (barrier synchronisation). OpenMP's parallel `for loops' allows for dynamic load balancing; (if a core has finished its work, it can be assigned more work in the meantime). Whereas the static load balancing in OpenMP is the same as the MPI implementation of the code. 
Traditionally, N-body simulations use `for loops' to calculate the distance between particles, determining the force. 
In contrast, the algorithms here (Serial, OpenMP, and MPI) use NumPy matrices (vectorisation) to calculate the distances, which is why dynamic load balancing couldn't be investigated. Vectorised force calculations are significantly faster than the `for loops'. 
Figure  6 compares the speedup ($T_{loop}/T_{vectorized}$) obtained by vectorising the force calculations for multiple model sizes ($n$), running on a single core ($p = 1$).
\vspace{-1em}
\section{4. Discussion \& improvements}
\subsection{4.1 Current algorithm }
Although both (shared and distributed memory) algorithms compute the forces in parallel, the system's state is updated on the master core. For larger model sizes, this update stage consumes quite a lot of time. A potential improvement would be to perform this computation in parallel as well. This would require two additional communication/synchronisation steps but could potentially speed up the algorithm. However, this approach would slow down the performance of the model size is too small for the same reasons discussed above. Compared to the point-to-point communication commands (MPI.Send and MPI.Recv), the collective MPI commands (MPI.Bcast and MPI.Gather) perform equally (if not worse for smaller models), which is why the collective MPI commands were not implemented in the MPI code. 

\begin{figure}[h]
\includegraphics[width=0.8\columnwidth]{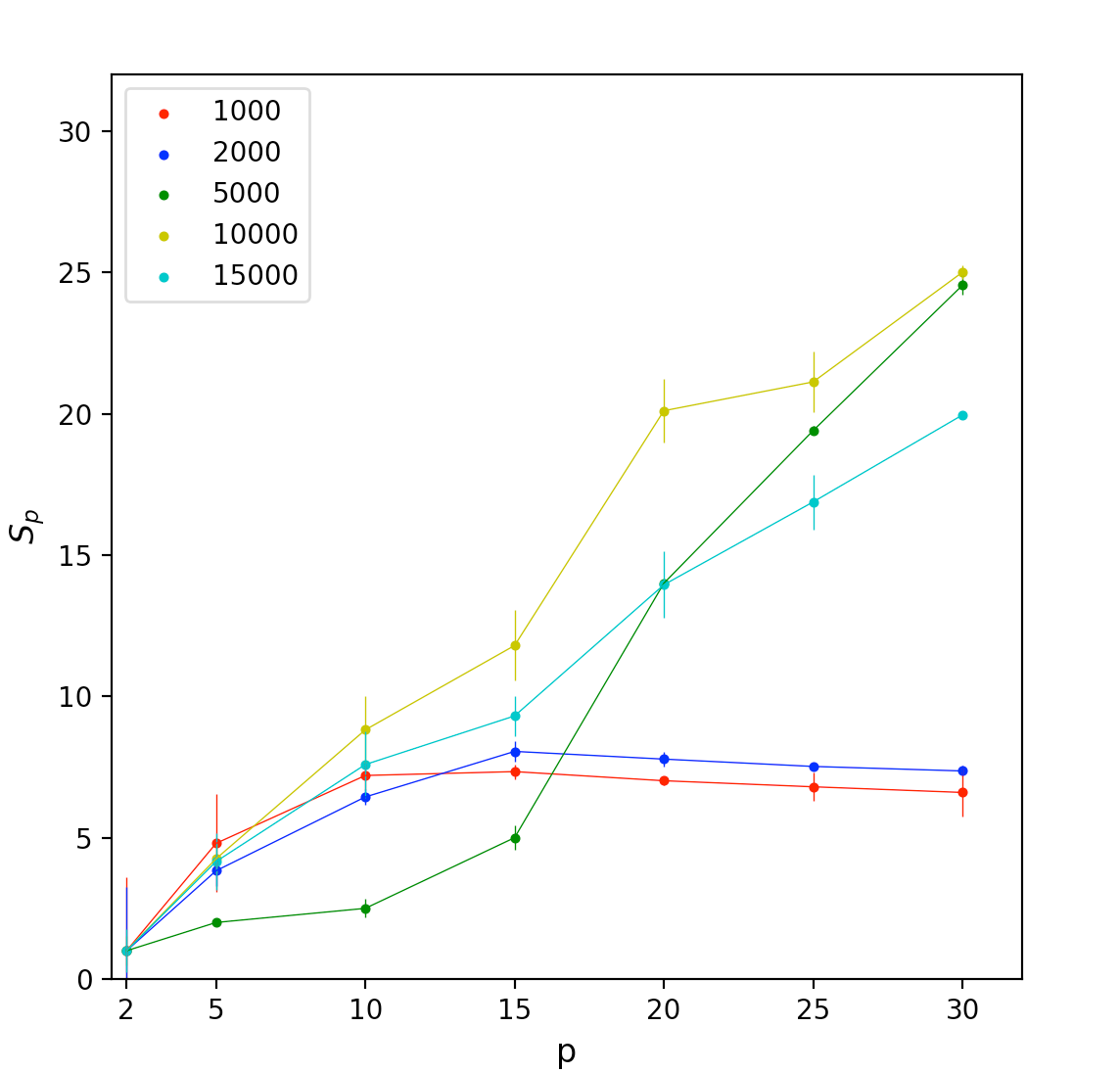}
\caption{Comparing how the MPI algorithm scales with the number of cores ($2\leq p \leq32$), for multiple model sizes ($5000 \leq n \leq 15000$). Each model size scales differently with the number of processes. Smaller models don't continue to linearly scale $p$ since it takes longer to distribute the data compared to performing the calculation itself. }
\end{figure}
\begin{figure}[h]
\includegraphics[width=0.8\columnwidth]{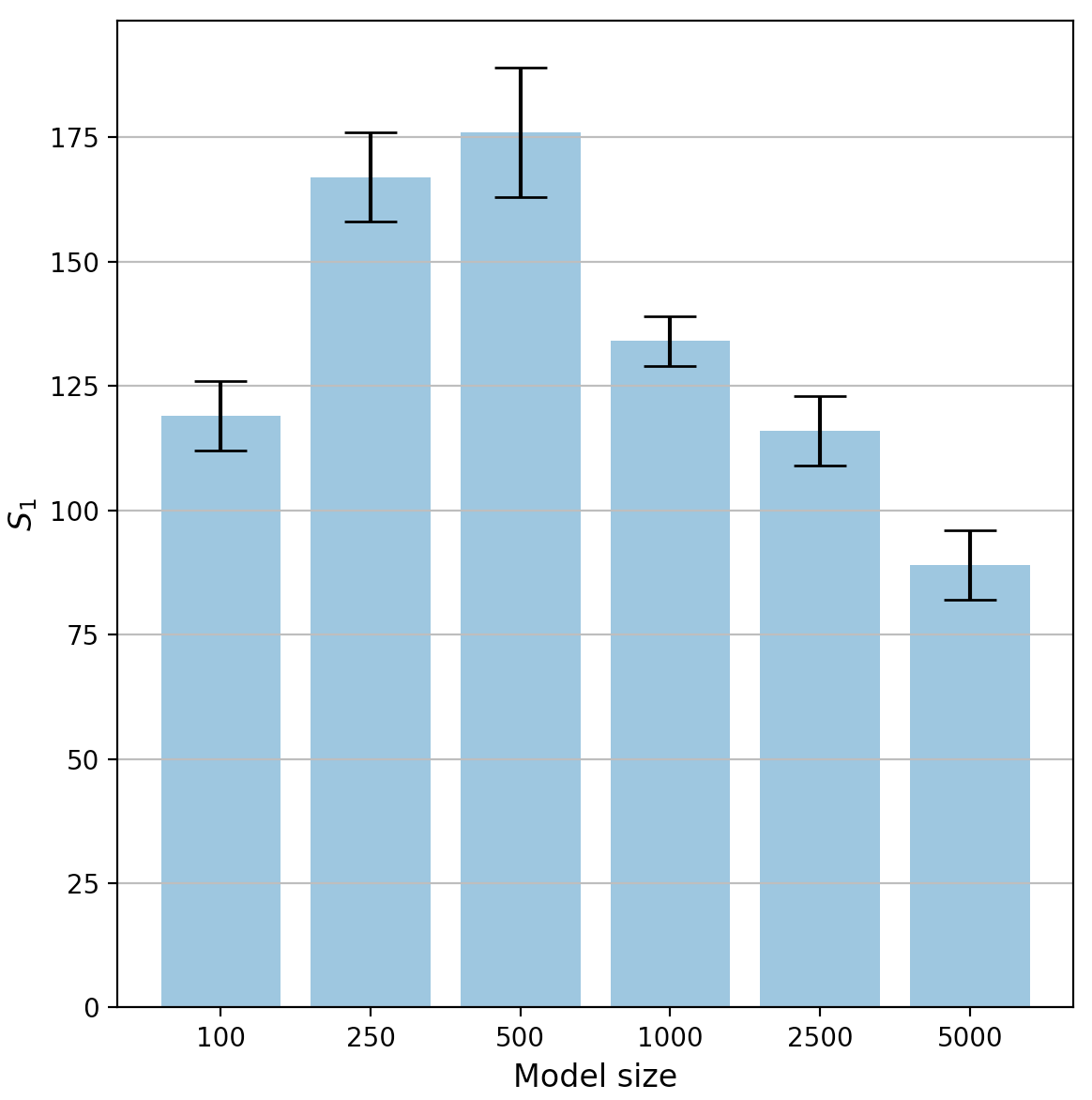}
\caption{The speedup obtained (serial algorithm) by replacing the `for loops' with vectorised operations (NumPy matrices) to calculate the distances (particle separation).
All the algorithms (Serial, OpenMP, and MPI) used vectorised force calculations.}
\end{figure}
 \vspace{-1em}
 
 \subsection{4.2 Memory interfaces}
 The MPI \& OpenMP memory architectures can be combined into a
 hybrid approach. This approach uses multiple processors (via MPI) and runs the OpenMP algorithm on every processor (cores in the same processor share memory). This Hybrid approach allows for dynamic load balancing to be scaled over a cluster of processors, improving the performance by up to 20\% \cite{9} compared to the state-of-the-art MPI algorithm. A new programming API known as High-Performance ParalleX (HPX) provides an asynchronous programming/memory environment which performs 30\% better than the best Hybrid model to date \cite{10}.

 \subsection{4.3 Computing architectures }
 A typical CPU contains $\mathcal{O}(10)$ powerful cores built on the MIMD (Multiple Instruction, Multiple Data) design. In contrast, the GPU contains $\mathcal{O}(1000)$ less powerful cores built on the SIMD (Single instruction, multiple data) design, making the GPU ideal for running multiple identical computations simultaneously (force calculations in this case). Studies indicate that GPU accelerated N-body simulations outperform (200\%) CPU exclusive systems \cite{11}. 
   
Recently, a new type of processor called the IPU (Intelligence Processing Unit) was explicitly launched and designed for machine learning applications. The IPU is built on the MIMD paradigm and contains 1216 individual processing units called tiles which can communicate with each other extremely quickly (45 TB/s, faster than a typical L1 cache). 
Each tile can run six threads simultaneously, resulting in a total of $6 \times 1216 = 7296$ threads which can be run in parallel on a single IPU processor \cite{12}. 
Since the primary source of overheads in the MPI algorithm is communication, the high-speed bandwidth offered by the IPU holds the potential to drastically speed up n-body simulations (via a distributed-memory approach). The particle physics department at the University of Bristol uses the IPU to study and develop High energy physics (HEP) algorithms for the LHCb/CMS experiment at CERN.  
The performance of the MPI algorithm (written in python) couldn't be evaluated on the IPU since the IPU's parallel programming API (Poplar) only supports C++ at the moment. However, the python version Poplar (supporting MPI) is expected to be released soon. 
   
\subsection{4.4 Alternate Methods }
 Finally, alternate methods for n-body simulations exist. The direct approach has a time complexity of $O(N^{2})$. The tree algorithm approximates the forces by dividing the simulation space into an octree. Force calculations are computed individually for particles below some threshold distance $\theta$ from the current particle. If a cluster of particles is beyond this threshold distance, then the cluster is considered a single particle, with its centre being the cluster's centre of mass. This reduces the total number of force calculations needed to be performed at every time step, resulting in time complexity of $O(N\log(N))$ \cite{13}. The Fast Multipole algorithm improves upon the tree algorithm by exploiting that nearby particle in a gravitational field experience similar acceleration from distant clusters of particles. The force contribution from a distant cluster only needs to be computed once, reducing the number of force calculations required, resulting in a time complexity of $O(N)$ \cite{14}.

\section{5. Conclusion}
The direct N-body algorithm was parallelised via two distinct memory systems; shared memory (OpenMP) and distributed memory (MPI). Both of these approaches were examined to evaluate the effectiveness of parallelization. Here, the speedup ($S_{p}$) was used as the performance metric. The OpenMP algorithm outperforms MPI algorithm since the shared memory architecture eliminates the need for the cores to communicate. However, the OpenMP approach is limited to a single processor (processors clusters don't share cache memory) which limits the maximum speedup obtainable on the number of cores a processor contains. In contrast, the distributive memory architecture of the MPI algorithm allows it to utilise multiple processors, which outperforms the OpenMP algorithm when evaluated on a large number of cores. The performance of each model size scales differently with the number of cores. Smaller models don't benefit much by the parallelization, since its often faster to perform the calculations on a single core than to distribute/perform the calculations in parallel. As the model size increases, the performance almost scales linearly with the number of cores $p$. Although due to Amdahl's law, the maximum speedup which can be obtained by an algorithm becomes independent of $p$ after ($p \geq p_{o}$) where $p_{o}$ is the optimal number of cores. This $p_{o}$ depends on the faction of the code which cannot be parallelized (sequential). Alternate methods of N-body simulations (Tree code, Fast Multipole) were briefly discussed, since they drastically reduce the time complexity of the simulation by incorporating some approximations.


\vspace{-2mm}

\section{References}
\vspace{-7mm}
\begin{thebibliography}{99}

\bibitem{1} \textit{1960: Siliconengine Metal-Oxide-Semiconductor (MOS) demonstrated}, computerhistory (2014), [ONLINE] Available at: www.computerhistory.org/siliconengine [Accessed 11 February 2021].

\bibitem{2} \textit{The Rise of MOS Technology \& The 6502}, commodore (2006), [ONLINE] Available at: www.commodore.ca/ [Accessed 11 February 2021].

\bibitem{3} \textit{Computer Architecture: A Quantitative Approach}, J. L. Hennessy, D. A. Patterson, Morgan Kaufmann publishers, (2011)

\bibitem{4} \textit{Parallel Computing: Review and Perspective}, Y. Li, Z. Zhang y, 2018 5th International Conference on Information Science and Control Engineering (ICISCE), (2018)

\bibitem{5} \textit{Amdahl's Law in the Multicore Era}, M. D. Hill et al, Computer vol. 41, (2008)

\bibitem{6} \textit{N-body simulations of gravitational dynamics}, W. Dehnen et al, Eur. Phys. J. Plus, (2011)

\bibitem{7} \textit{N Towards optimal softening in three-dimensional N-body codes}, W. Dehnen, Royal Astronomical Society, (2001)

\bibitem{8} \textit{Time stepping N-body simulations}, T. Quinnl et al, 	arXiv: Astrophysics, (1997)

\bibitem{9} \textit{Hybrid MPI-OpenMP Paradigm on SMP clusters: MPEG-2 Encoder and n-body Simulation}, K. Yamazaki et al, 	arXiv: Distributed, Parallel, and Cluster Computing, (2012)

\bibitem{10} \textit{A Massively Parallel Distributed N-body Application Implemented with HPX}, Z. Khatami et al, 	7th Workshop on Latest Advances in Scalable Algorithms for Large-Scale Systems, (2016)

\bibitem{11} \textit{Toward efficient GPU-accelerated N-body simulations}, M.J. Stock, 46th AIAA Aerospace Sciences Meeting and Exhibit, (2012)

\bibitem{12} \textit{Dissecting the Graphcore IPU Architecture via Microbenchmarking}, Z. Jia, B. Tillman et al, arXiv: Distributed, Parallel, and Cluster Computing, (2019)

\bibitem{13} \textit{A hierarchical O(N log N) force-calculation algorithm}, J. Barnes, P. Hut, Nature 324, (1986)

\bibitem{14} \textit{A Fast Algorithm for Particle Simulations}, L. Greengard, V. Rokhlin, Journal of Computational Physics, (1987)

\end{thebibliography}
\end{document}